\documentclass[twocolumn,showpacs,amsmath,amssymb]{revtex4}
\usepackage{graphicx}
\usepackage{dcolumn}
\usepackage{bm}
\newcommand{\beq}{\begin{equation}}
\newcommand{\eeq}{\end{equation}}
\newcommand{\bea}{\begin{eqnarray}}
\newcommand{\eea}{\end{eqnarray}}
\newcommand{\eps}{\varepsilon}

\begin{document}

\author{E. E. Saperstein}
\affiliation{Kurchatov Institute, 123182 Moscow, Russia}
\affiliation{National Research Nuclear University MEPhI, 115409
Moscow, Russia}

\author{M. Baldo}
\affiliation{INFN, Sezione di Catania, 64 Via S.-Sofia, I-95125 Catania, Italy}

\author{N. V. Gnezdilov}
\affiliation{Instituut-Lorentz, Universiteit Leiden, P.O. Box 9506, 2300 RA Leiden, The Netherlands}

\author{S. V. Tolokonnikov}
\affiliation{Kurchatov Institute, 123182 Moscow, Russia}
\affiliation{Moscow Institute of Physics and Technology, 141700
Dolgoprudny, Russia.}

\title{Phonon contributions to ab initio double  mass differences of magic nuclei}


\begin{abstract}Odd-even double mass differences (DMD) of magic nuclei are found within the approach
starting from the free $NN$ interaction with account for particle-phonon coupling (PC) effects.  We
consider three PC effects: the phonon induced effective interaction,  the renormalization of the
``ends'' due to the $Z$-factor corresponding to the pole PC contribution to the nucleon mass operator
and the change of the single-particle energies. The perturbation theory in $g^2_L$, where $g_L$ is the
vertex of the $L$-phonon creation, is used for PC calculations.
   PC  corrections to single-particle energies are found self-consistently
with an approximate account for the tadpole diagram.   Results for magic $^{40,48}$Ca, $^{56,78}$Ni,
$^{100,132}$Sn  and $^{208}$Pb nuclei are presented. For lighter part of this set of nuclei, from
$^{40}$Ca till $^{56}$Ni, the cases divide approximately in half between those where the PC
corrections to DMD values make agreement with the data better and the ones with the opposite result.
In the major part of the cases of worsening of description of DMD, a poor applicability of the
perturbation theory for the induced interaction is the most probable reason of the phenomenon. For
intermediate nuclei, $^{78}$Ni and  $^{100}$Sn, there is no sufficiently accurate data on masses of
nuclei necessary for finding DMD values. Finally, for heavier nuclei, $^{132}$Sn and  $^{208}$Pb, PC
corrections always make agreement with the experiment better.
\end{abstract}

\maketitle

\section{Introduction} Last decade, an  {\it ab initio}  approach to the nuclear pairing
problem starting from free $NN$ potential was successfully developed. The first work of the Milan group on this subject
\cite{milan1} played the key role showing that the solution of the BCS gap equation for the
nucleus $^{120}$Sn with the realistic  Argonne v$_{14}$ potential and the Saxon-Woods Shell-Model basis with the bare
neutron mass $m^*=m$  gives a reasonable result, $\Delta_{\rm
BCS}=2.2$ MeV. Although it is bigger of the experimental one,
$\Delta_{\rm exp}\simeq 1.3$ MeV, the difference is not so dramatic
leaving a hope to achieve a good agreement by developing
corrections to the scheme. In Refs. \cite{milan2,milan3} the basis
was enlarged from  $E_{\rm max}=600$ MeV in \cite{milan1} to $E_{\rm max}=800$ MeV, and
the effective mass $m^*\neq m$ was introduced into the gap equation.
The new basis was calculated within the Skyrme--Hartree--Fock
method with the Sly4 force \cite{SLy4}, that makes the effective
mass $m^*(r)$ coordinate dependent and essentially different from
the bare one. E.g., in nuclear matter the Sly4 effective mass is equal to $m^*=0.7 m$.
So small value of the effective mass lead in to a strong suppression of the
gap value to  $\Delta_{\rm BCS}=0.7$ MeV in \cite{milan2} or $\Delta_{\rm
BCS}=1.04$ MeV  in \cite{milan3}. In both cases, the too small
value of the gap was explained by invoking various many-body
corrections to the BCS approximation. The main correction is due to
the exchange of low-lying surface vibrations (``phonons''),
contributing to the gap about 0.7 MeV \cite{milan2}, so that the
sum  turns out to be $\Delta=1.4$ MeV very close to the experimental
value. In Ref. \cite{milan3}, the contribution of the induced
interaction caused by exchange of the high-lying in-volume
excitations was added either, the sum again is equal to
$\Delta\simeq 1.4$ MeV. Thus, the calculations of Refs.
\cite{milan2,milan3} showed that the effects of the effective mass $m^*\neq m$ and of
many-body corrections to the BCS theory are necessary  to explain
the difference of ($\Delta_{\rm BCS}-\Delta_{\rm exp}$). In
addition, their contributions are of different sign and partially compensate each
other. Unfortunately, both effects contain large uncertainties. This
point was discussed in detail in Refs. \cite{Bald1,BCS50}.

A bit later, Duguet and Losinsky \cite{Dug1} made an important step in the
problem by solving the {\it ab initio} BCS gap equation for a lot of
nuclei on the same footing. It should be noticed that the main
difficulty of the direct method to solve the nuclear pairing problem
comes from the rather slow convergence of the sums over intermediate
states $\lambda$ in the gap equation because of the short-range of
the free $NN$-force. This, evidently, was the main reason why the Milan group limited
their investigations \cite{milan1,milan2,milan3} to only the nucleus $^{120}$Sn. To avoid the slow convergence, the authors of
Refs. \cite{Dug1,Dug2} used the ``low-k'' force ${\cal V}_{\rm low-k}$ \cite{Kuo,Kuo-Br} which is in fact very soft. ${\cal V}_{\rm low-k}$ is defined
in such a way that it describes correctly the $NN$-scattering phase
shifts at momenta $k{<}\Lambda$, where $\Lambda$  is a parameter
corresponding to
 the limiting energy  $\simeq 300\;$MeV.  The force ${\cal V}_{\rm low-k}$
vanishes  for $k{>}\Lambda$, so that in the gap equation one can
restrict the energy range
 to $E_{\max} {\simeq} 300\;$MeV. In addition, a separable version of this force
was constructed that made it possible to calculate neutron and
proton pairing gaps for a lot of nuclei. Usually the low-k force is
found starting from some realistic $NN$-potential ${\cal V}$ with
the help of the Renormalization Group method, and the result does
not practically depend on the particular choice  of ${\cal V}$
\cite{Kuo}. In addition, in Ref. \cite{Dug1} ${\cal V}_{\rm low-k}$
was found starting from the Argonne potential v$_{18}$, that is
different only a little from Argonne v$_{14}$, used in Ref.
\cite{milan3}. Finally, in Ref. \cite{Dug1} the same SLy4
self-consistent basis was used as in Ref. \cite{milan3}. Thus, the
inputs of the two calculations look very similar, but the results
turned out to be strongly different. In fact, in Ref. \cite{Dug1}
the value $\Delta_{\rm BCS}\simeq 1.6\;$MeV was obtained for the
same nucleus $^{120}$Sn which is already bigger than the
experimental one by $\simeq 0.3\;$MeV. In Refs. \cite{Bald1,Pankr1}
the reasons of these contradictions were analyzed.  It turned out
that these two calculations differ in the way they take
into account the effective mass. It implies that the gap $\Delta$
depends not only on the value of the effective mass at the Fermi
surface, as it follows from the well-known BCS exponential formula for the
gap, but also on the behavior of the function $m^*(k)$ in a wide
momentum range. However, this quantity is not known sufficiently  well.
An additional problem was specified in Ref. \cite{Dug3} where it was
found that the inclusion of the {\it ab initio} 3-body force following
from the chiral theory \cite{Epel}  suppresses the gap values much lower than the
experimental ones.

To avoid all these uncertainties, a semi-microscopic model for nuclear
pairing was suggested by the Moscow-Catania group
\cite{Pankr1,Pankr2,Sap1}. It starts from the {\it ab initio} BCS
gap equation with the Argonne $NN$-potential v$_{18}$ treated with the
two-step method. The complete Hilbert space of the problem is
split into the model subspace of low-energy states and the
complementary one. The gap equation is solved in the model
space with the effective pairing interaction (EPI) ${\cal V}_{\rm eff}$ which is found
in the complementary subspace in terms of the initial $NN$-potential
${\cal V}$. The self-consistent basis of the energy density functional (EDF) by Fayans et al.
\cite{Fay1,Fay4,Fay5,Fay} was used which is characterized with the bare mass $m^*=m$. The set DF3 of the EDF parameters \cite{Fay4,Fay}
was chosen or its modified version DF3-a \cite{Tol-Sap}. The modification concerns the spin-orbit and effective tensor terms of the Fayans EDF.
This is not much important for the pairing problem \cite{Pankr2,Sap1} but there is a noticeable difference between these two EDFs, in favor of DF3-a, in some other problems, e. g. in calculating characteristics of the first $2^+$-states in semi-magic nuclei \cite{BE2}.

A new version of the
local approximation, the so-called Local Potential Approximation
(LPA) \cite{Rep}, is used in the complementary subspace to simplify calculations.  This {\it
ab-initio} term of ${\cal V}_{\rm eff}$ is supplemented by a small addendum
proportional to the phenomenological parameter $\gamma$ that should
hopefully embody all corrections to the simplest BCS scheme with
$m^*=m$. Smallness of the correction term is demonstrated in Fig. 1
where a localized ``Fermi average'' form of ${\cal V}_{\rm eff}$ is displayed
without ($\gamma=0$) and with ($\gamma=0.06$) the phenomenological
correction. Non-negligible effect of so small change of ${\cal V}_{\rm eff}$ to
the gap value is owing  to the exponential dependence of the gap on the strength
of pairing force mentioned above.

\begin{figure}
\vspace{-5mm} \centerline {\includegraphics [width=80mm]{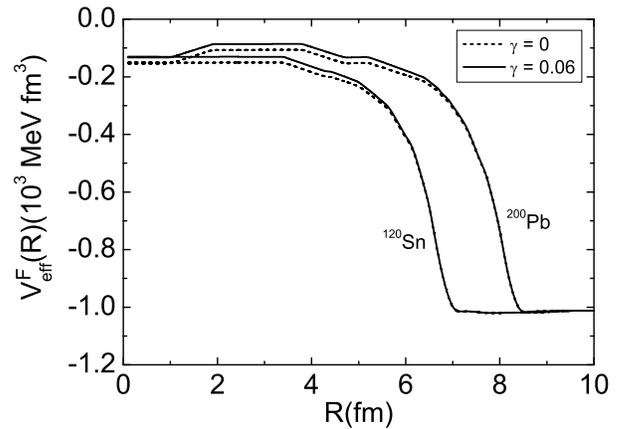}} \vspace{-2mm}
\caption{The Fermi average effective pairing interaction ${\cal
V}^{\rm F}_{\rm eff}(R)$ for $^{120}$Sn and $^{200}$Pb nuclei}
\end{figure}

The ``experimental'' gap value $\Delta_{\rm exp}$ for semi-magic nuclei is
usually identified with
a half of one of the following odd-even double mass differences (DMD):
\beq D_{2n}^+(N,Z) =  M(N+2,Z)+ M(N,Z)-2M(N+1,Z),\label{d2npl} \eeq
\beq D_{2n}^-(N,Z) =  -M(N-2,Z)- M(N,Z)+2M(N-1,Z),\label{d2nmi}\eeq
\beq D_{2p}^+(N,Z) =  M(N,Z+2)+ M(N,Z)-2M(N,Z+1),\label{d2ppl} \eeq
\beq D_{2p}^-(N,Z) = -M(N,Z-2)-M(N,Z)+ 2M(N,Z-1).\label{d2pmi}\eeq

The accuracy of such a prescription  was estimated in \cite{Pankr2} as
${\simeq}0.1\div0.2\;$MeV. Approximately the same accuracy holds for the ``developed pairing'' approximation
in the gap equation, with conservation of the particle number only on average \cite{part-numb},
used in all References on the pairing problem cited above.

There is one more physical quantity in semi-magic nuclei which can be evaluated in terms of
the same effective interaction as the pairing gap. This is the set of the same double odd-even
mass differences (\ref{d2npl})--(\ref{d2pmi}), but for the non-superfluid subsystems. Now
$N$ is magic and $Z$ arbitrary in Eqs. (\ref{d2npl}),(\ref{d2nmi}) and {\it vice versa} in Eqs. (\ref{d2ppl}),(\ref{d2pmi}).
In non-superfluid nuclei, the mass differences, Eqs.
(\ref{d2npl}),(\ref{d2nmi}), coincide with poles in the total energy $E$ plane of the two-particle
Green function $K(1,2,3,4)$ for  normal systems  \cite{AB} in the $nn$-channel,
and Eqs. (\ref{d2ppl}),(\ref{d2pmi}), in the $pp$-channel. The equation for $K$ in the channel $S=0,L=0$
could be expressed in terms of
the same EPI ${\cal V}_{\rm eff}$   as the pairing gap.
This point was marked  in the old paper \cite{Sap-Tr}, where these differences for double-magic nuclei were analyzed
within the theory of finite Fermi systems (TFFS) \cite{AB}. In that article,
the density dependent EPI was introduced for the first time and arguments were found in favor of
of the surface dominance in this interaction.

It is worth to stress that this calculation of mass difference for the non-superfluid subsystem within the mean field theory is a more
rigorous operation than its identification with the double gap $\Delta$ of the BCS scheme in the superfluid one.
The first such calculations with the use of the semi-microscopic model for the effective
pairing interaction with the same value $\gamma=0.06$ of the phenomenological parameter of the model,
found previously in the pairing problem, were carried out recently for several semi-magic chains \cite{Gnezd1,Gnezd2,Gnezd3}.

In this work, we analyze corrections to the mean field theory of double odd-even mass differences due to particle-phonon coupling (PC)
effects.  Three PC effects are taken into
account, the phonon induced interaction,  the renormalization of the ``ends'' due to the $Z$-factor
corresponding to the pole PC contribution to the nucleon mass operator and the change of the
single-particle energies.  In the latter case, the non-pole (so-called ``tadpole'') diagram for the mass operator is taken into account
in addition to the usual pole one. We limit ourselves with the double-magic nuclei where the perturbation theory on the PC vertex $g_L$ is
usually valid.

\section{Brief formalism}

\subsection{The semi-microscopic model of the effective pairing interaction}
To begin with, we describe briefly the semi-microscopic model of the EPI which will be used for finding the double odd-even mass differences in non-superfluid nuclei.
The general many-body form of the equation for the pairing gap is as
follows \cite{AB},
\beq \Delta=  {\cal U} G G^s
\Delta, \label{del} \eeq where   ${\cal U}$ is the $NN$-interaction block irreducible
in the two-particle channel, and
 $G$  ($G^s$) is the one-particle Green function without (with)
 pairing. We consider the singlet, $S=0$ and $L=0$, pairing only. The isospin indices are omitted for brevity. A symbolic multiplication denotes the integration over
energy and intermediate coordinates and summation over spin
variables as well.  In the Brueckner theory, first, the block ${\cal U}$ should be replaced with the free $NN$-potential ${\cal V}$,
which does not depend on the energy. Second, simple quasi-particle Green functions $G$ and $G^s$ are used, i.e. those without PC corrections and so on.
In the result, Eq. (\ref{del}) coincides with the one of the BCS approximation and can be reduced to the form usual for the Bogolyubov method,
\beq \Delta = - {\cal V} \varkappa\,, \label{delkap}\eeq where \beq\varkappa=\int \frac {d\eps}{2\pi i}G G^s\Delta \label{defkap}\eeq is the
anomalous density matrix.

As it was discussed in Introduction, Eq. (\ref{del}) converges very slowly due to
the short-range character of $NN$ potential. To overcome this problem, a two-step renormalization
method of solving the gap equation in nuclei was used in
Refs. \cite{Pankr1,Pankr2,Sap1}.   The complete Hilbert space of the
pairing problem $S$ is split in the model subspace $S_0$, that includes
the single-particle states with energies less than a separation
energy $E_0$, and the complementary one, $S'$. The gap equation is
solved in the model space: \beq \Delta= {\cal V}_{\rm eff} G
G^s \Delta|_{S_0}, \label{del0} \eeq with the
EPI ${\cal V}_{\rm eff}$ instead of the block
${\cal V}$ in the BCS version of the original gap equation
(\ref{del}). It obeys the Bethe--Goldstone type equation in the
subsidiary space, \beq {\cal V}_{\rm eff} = {\cal V}  + {\cal
V}  G  G {\cal V}_{\rm eff}|_{S'}. \label{Vef} \eeq In
this equation, the pairing effects can be neglected provided the
model space is sufficiently large, $E_0\gg \Delta$. That is why we
replaced the Green function $G^s$ for the superfluid system
with its counterpart $G$ for the normal system.  The problem of slow convergence has passed now to
 Eq. (\ref{Vef}) for the EPI  ${\cal V}_{\rm eff}({\bf r}_1,{\bf r}_2,{\bf r}_3,{\bf r}_4)$. To solve it,
the LPA method is used as it was discussed in the Introduction. It turned out \cite{Rep} that, with a very high accuracy,
at each value of the average c.m. coordinate ${\bf R}=({\bf r}_1 +
{\bf r}_2 +{\bf r}_3 +{\bf r}_4)/4$, one can use in Eq. (\ref{Vef})
the formulae  of the infinite system embedded into a constant
potential well $U=U({\bf R})$. This significantly simplifies the
equation for ${\cal V}_{\rm eff}$, in comparison with the initial equation for
$\Delta$. As a result, the subspace $S'$ can be chosen as large as
necessary to achieve the convergence. Accuracy of  LPA depends on
the separation energy $E_0$. For finite nuclei, the value of
$E_0{=}40\;$MeV guarantees the accuracy better than 0.01 MeV for the
gap $\Delta$.

To avoid uncertainties of explicit consideration of corrections to
the BCS scheme discussed above, the semi-microscopic model was
suggested in Refs.\cite{Pankr1,Pankr2,Sap1}. In this model, a small
phenomenological addendum to  the EPI is
introduced which embodies in an effective way all these corrections. The
simplest ansatz for it was used: \bea {\cal V}_{\rm
eff}({\bf r}_1,{\bf r}_2,{\bf r}_3,{\bf r}_4) = V^{\rm
BCS}_{\rm eff}({\bf r}_1,{\bf r}_2,{\bf r}_3,{\bf r}_4) + \nonumber \\
\gamma C_0 \frac {\rho(r_1)}{\bar{\rho}(0)}
\prod_{i=2}^4\delta ({\bf r}_1 - {\bf r}_i). \label{Vef1} \eea Here
$\rho(r)$ is the density of nucleons of the kind under
consideration, and $\gamma$ are dimensionless
phenomenological parameters. To avoid any influence of the shell
fluctuations in the value of ${\rho}(0)$, the average central
density ${\bar{\rho}(0)}$ is used in the denominator of the
additional term. It is averaged over the interval of $r{<}2\;$fm.
The first, {\it ab initio}, term in the r.h.s. of Eq. (\ref{Vef1})
is the solution of  Eq. (\ref{Vef}) in the framework of the LPA
method described above, with $m^*{=}m$ in the subspace $S'$.

\subsection{Double mass differences in magic nuclei}
As it is was discussed in Introduction, the double odd-even mass differences (\ref{d2npl})--(\ref{d2pmi}) in non-superfluid nuclei can be expressed
in terms of the same EPI (\ref{Vef}) as the gap (\ref{del0}).
To derive the equation for this quantity, it is convenient to start from the Lehmann expansion  for the two-particle Green function $K$ in a non-superfluid system.
In the single-particle wave functions $|1\rangle{=}|n_1,l_1,j_1,m_1\rangle$ representation, it  reads \cite{AB}:
\beq K_{12}^{34}(E)=\sum_s
\frac {\chi^s_{12}\chi^{s+}_{34}} {E-E_s^{+,-} \pm i\gamma},
\label{Lem}\eeq where $E$ is the total energy in the two-particle channel and
$E_s^{+,-}$ denote the eigen-energies of  nuclei with
 two particles and two holes, respectively, added to the original nucleus. They are often interpreted as the ``pair
vibrations'' \cite{BM2}.  Instead of the Green function $K$, it is convenient
to use the two-particle interaction amplitude
$\Gamma$: \beq K = K_0 + K_0 \Gamma K_0, \label{gam}\eeq where
$K_0=GG$. The amplitude
$\Gamma$ obeys the following equation \cite{AB}: \beq \Gamma = {\cal U}+{\cal U} GG
\Gamma, \label{eqgam}\eeq where ${\cal U}$ is the same irreducible interaction block as in Eq. (\ref{del}).
Again, within the Brueckner theory,  the block  ${\cal
U}$ should be replaced with the realistic $NN$-potential ${\cal V}$
which does not depend on the energy. Then the integration over the relative energy can be readily carried
out in Eq. (\ref{eqgam}): \beq
 A_{12} {=}  \int \frac {d\eps}{2\pi
i}G_1\left(\frac E 2 {+}\eps \right) G_2\left(\frac E 2 {-}\eps
\right)
 {=}\frac {1{-}n_1{-}n_2}
{E{-}\eps_1{-}\eps_2}, \label{Alam} \eeq where $\eps_{1,2}$
are the single-particle energies and $n_{1,2}{=}(0;1)$, the corresponding occupation numbers.
As the result, we obtain: \beq \Gamma = {\cal V}+{\cal V} A
\Gamma. \label{eqgam1}\eeq

In vicinity of a pole  $E{=}E_s$, one gets \beq  \Gamma = \frac {d_s d_s^+}{E-E_s},  \label{Gampole}\eeq
where $d_s^+\; (d_s)$ are vertices of creation (anihillation) of the two-particle state $|s\rangle$, the non-homogeneous Eq. (\ref{eqgam1}) reduces to the homogeneous one,
\beq \Gamma = {\cal V} A
\Gamma, \label{eqBS}\eeq which is, in fact, the in-medium Bethe-Salpeter equation, or equivalently \beq d_s = {\cal V} A
d_s, \label{eqgs}\eeq.

It is more convenient to transform this equation to the one for the eigenfunctions  $\chi^s=Ad_s$: \beq (E_s-\eps_1-\eps_2)
\chi^s_{12}=(1-n_1-n_2) \sum_{34}{\cal V}_{12}^{34}
\chi^s_{34}\label{eqchi}.\eeq It is different from the Shr\"{o}dinger equation for
two interacting particles in an external field only with the factor
$(1-n_1-n_2)$ which reflects the many-body character of the problem, in particular, the Pauli principle.
As in the pairing problem, the angular momenta of two-particle states  $|12\rangle$, $|34\rangle$ are coupled to
the total angular momentum  $I{=}0$ ($S{=}0, L{=}0$).

The direct solution of this equation is  complicated by the same reasons as for the {\it ab initio} BCS
gap equation described above. The same two-step method is used in combination with LPA to overcome this difficulty. The usual renormalization
of Eq. (\ref{eqchi}) transforms it into the analogous equation in the model space:\beq (E_s{-}\eps_1{-}\eps_2)
\chi^s_{12}{=}(1{-}n_1{-}n_2) {\sum_{34}}^0 \left({\cal V}_{\rm eff}\right)_{12}^{34}\; \chi^s_{34},\label{eqchi0}\eeq where
the effective interaction ${\cal V}_{\rm eff}$ coincides with that
of pairing problem, Eq. (\ref{Vef}), provided the same value of the separation energy $E_0$ is used.
It agrees with the well-known theorem by Thouless  \cite{Thouless} stating that the gap equation reduces to the in-medium
Bethe-Salpeter equation provided the gap $\Delta$ vanishes.
The next step consists in the use of the ansatz (\ref{Vef1})
to take into account corrections to the Brueckner theory with a phenomenological addendum
($ \sim \gamma$).

The double mass differences (\ref{d2npl})--(\ref{d2pmi}) are identified with the two first solutions
 $E_s^{+,-}$ of Eq. (\ref{eqchi0}), corresponding to the addition of two particles (holes) to the
 magic core into the state $\eps_1=\eps_2=\mu^{+,-}$, where the chemical potentials
$\mu^{+,-}$ are defined in a usual way as mass differences, e.g., $\mu_p^+=E_{\rm B}(N,Z+1)-E_{\rm B}(N,Z)$. Then,
the energy difference in the left-hand side of  Eq. (\ref{eqchi0}) is equal directly the quantity we need:
$E_s^{+,-}-2\mu^{+,-}=D^{+,-}$.

In Refs. \cite{Gnezd1,Gnezd2,Gnezd3} this scheme of finding the DMD $D^{+,-}$ in non-superfluid systems
within the semi-microscopic model was used for several chains of semi-magic nuclei. The proton subsystem should be considered for isotopic chains with magic $Z$ value and the neutron one, for isotonic chains where $N$ is magic. Reasonable results were obtained with the same value $\gamma{=}0.06$ which was previously found for the pairing gap \cite{Pankr2}. In this work, we study in an explicit form the PC effects in this problem and analyze possibility of modifying the optimal value of the parameter $\gamma$.  This can be expected, since the PC effect was implicitly included in $\gamma$.

\subsection{Particle-phonon coupling contributions to double mass differences}
Introducing of the PC corrections to Eq. (\ref{eqchi0}) consists, first, of the change of $\eps_{\lambda}$ on the l.h.s. to $\widetilde{\eps}_{\lambda}{=}\eps_{\lambda}{+}\delta \eps_{\lambda}^{\rm PC}$ and, second, a similar change of the ${\cal V}_{\rm eff}$ quantity on the r.h.s., to $\widetilde{\cal V}_{\rm eff}$, with the same meaning of the ``tilde'' symbol. Let us write down this PC corrected equation explicitly:
\beq (E_s{-}\widetilde{\eps}_1{-}\widetilde{\eps}_2)
\chi^s_{12}{=}(1{-}n_1{-}n_2) {\sum_{34}}^0 \left(\widetilde{\cal V}_{\rm eff}\right)_{12}^{34}\; \chi^s_{34},\label{eqPCchi0}\eeq

Let us begin with more transparent part of the problem concerning the single-particle energies. We follow here the method developed in \cite{levels}.
Note also that recently PC corrections to the single-particle energies within different self-consistent approaches were studied in Refs.
\cite{Litv-Ring,Bort,Dobaczewski,Baldo-PC}.

To find the single-particle energies with account for the PC effects, we solve the following equation: \beq \left(\eps-H_0
-\delta \Sigma^{\rm PC}(\eps) \right) \phi =0, \label{sp-eq}\eeq where
$H_0$ is the quasiparticle Hamiltonian with the spectrum
$\eps_{\lambda}^{(0)}$ and $\delta \Sigma^{\rm PC}$ is the PC
correction to the quasiparticle mass operator. After expanding this
term in the vicinity of $\eps=\eps_{\lambda}^{(0)}$ one finds \beq
\eps_{\lambda}=\eps_{\lambda}^{(0)} + Z_{\lambda}^{\rm PC} \delta
\Sigma^{\rm PC}_{\lambda\lambda}(\eps_{\lambda}^{(0)})
,\label{eps-PC}\eeq with obvious notation. Here $Z^{\rm PC}$ denotes
the $Z$-factor due to the PC effects,

\beq Z_{\lambda}^{\rm PC} =\left({1- \left(\frac {\partial
\delta \Sigma^{\rm PC}(\eps)} {\partial \eps}\right)_{\eps=\eps_{\lambda}^{(0)}}}\right)^{-1}. \label{Z-fac}\eeq

Expression (\ref{eps-PC}) corresponds to the perturbation
theory in the $\delta \Sigma$ operator with respect to $H_0$. In this
article, we limit ourselves to magic nuclei where the so-called
$g_L^2$-approximation, $g_L$ being the $L$-phonon creation
amplitude, is, as a rule, valid. It is worth mentioning that Eq.
(\ref{eps-PC}) is more general, including, say, $g_L^4$ terms. In the case when several $L$-phonons are taken into account,
the total PC variation of the mass operator in Eqs. (\ref{sp-eq})--(\ref{Z-fac}) is just the sum:
\beq \delta \Sigma^{\rm PC} = \sum_L \Sigma^{\rm PC}_L . \label{sum-L}\eeq

\begin{figure}
\centerline {\includegraphics [width=60mm]{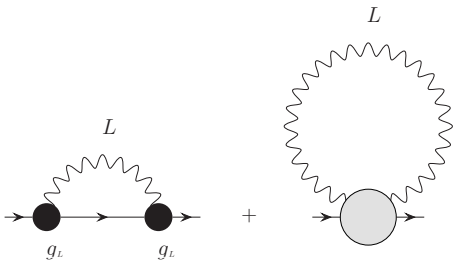}} \caption{PC corrections
to the mass operator. The gray blob denotes the ``tadpole'' term.} \label{fig:PC3}
\end{figure}

The diagrams for the $\delta \Sigma^{\rm PC}_L$ operator within the $g_L^2$-approximation are displayed in Fig. 2.
The first one is the usual pole diagram, with obvious notation, whereas the second, ``tadpole'' diagram represents the sum
of all non-pole diagrams of the $g_L^2$ order. For the pole term we are here neglecting the correction due to the one ``bubble'' diagram \cite{Baldo-PC}.
This can be justified provided only collective phonons are included. In the case of phonons of smaller collectivity, as e.g. positive parity states in
$^{208}$Pb, this correction could be important.

In the obvious symbolic notation, the pole diagram corresponds to
$\delta\Sigma^{\rm pole}=(g_L,D_LGg_L)$  where $D_L(\omega)$ is the
phonon $D$-function.  Explicit expression for the pole term is well known,  but we present it for completeness:
\bea \delta\Sigma^{\rm pole}_{\lambda\lambda}(\epsilon)&=&\sum_{\lambda_1\,M}
|\langle\lambda_1|g_{LM}|\lambda\rangle|^2 \nonumber\\
&\times&\left(\frac{n_{\lambda_1}}{\eps+\omega_L-
\eps_{\lambda_1}}+\frac{1-n_{\lambda_1}}{\eps-\omega_L -\eps_{\lambda_1}}\right), \label{dSig2} \eea
where $\omega_L$ is the excitation energy of the $L$-phonon. The $Z^{\rm PC}$-factor (\ref{Z-fac}) can be easily found from (\ref{dSig2}) by finding
the derivative over the energy $\eps$.

The vertex $g_L$  obeys the TFFS RPA-like equation \cite{AB}, \beq { g_L}(\omega)={{\cal F}} A_{\rm ph}(\omega) {g_L}(\omega), \label{g_L} \eeq
where ${\cal F}$ is the Landau--Migdal (LM) interaction amplitude, and $ A_{\rm ph}(\omega)=\int G \left(\eps + \omega/ 2 \right)
G \left(\eps -\omega/ 2 \right)d \eps/(2 \pi i)$ is the particle-hole propagator.  It
is normalized as follows \cite{AB}:
\beq\label{norm}
\left(g_L^+ \frac {d A_{\rm ph}}{d\omega} g_L \right)_{\omega=\omega_L}=-1,\eeq with obvious notation.

We use the self-consistent scheme to solve Eq. (\ref{g_L}) within the EDF method with the energy functional
\beq E_0=\int {\cal E}[\rho_n({\bf r}),\rho_p({\bf r})]d^3r,\label{E0} \eeq  where ${\cal E}$  is the energy density. In this approach, the LM amplitude
is found as the second variation derivative,
\beq {\cal F}_{\tau \tau'}=\frac {\delta^2 {E_0}}{\delta \rho_\tau \delta \rho_{\tau'}}, \label{LM} \eeq $\tau{=}n,p$ being the isotopic index. The Fayans
EDF we deal depends not only on the normal densities $\rho_\tau$ but on their anomalous counterparts  $\nu_\tau$  as well. However, we deal now with magic
nuclei where the anomalous densities vanish and we use therefore a simplified form (\ref{E0}) for $E_0$.

\begin{figure}
\centerline {\includegraphics [width=80mm]{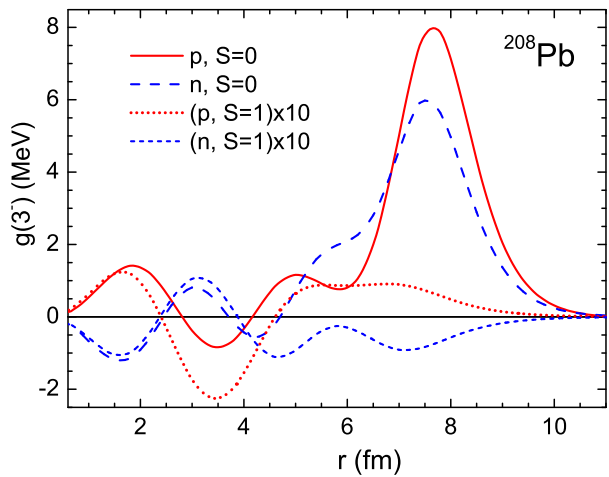}} \caption{ The vertex $g_L$ for the
$3^-_1$ state in $^{208}$Pb.} \label{fig:Pb3-}
\end{figure}

All the low-lying phonons we consider have  natural parity. In
this case, the vertex $g_L$ possesses  even $T$-parity. It is a
sum of two components with spins $S=0$ and $S=1$, respectively, \beq
g_L= g_{L0}(r) T_{LL0}({\bf n,\alpha}) +  g_{L1}(r)
T_{LL1}({\bf n,\alpha}), \label{gLS01} \eeq where $T_{JLS}$ stand
for the usual spin-angular tensor operators \cite{BM1}. The
operators $T_{LL0}$ and $T_{LL1}$ have  opposite $T$-parities, hence
the spin component should be the odd function of the excitation
energy, $g_{L1}\propto \omega_L$. This is the main reason why the $S=0$ component dominates
in such states. It is demonstrated in Fig. \ref{fig:Pb3-} for $3^-_1$ state in $^{208}$Pb, where the $S=1$ components
are multiplied by the factor 10 to be distinguishable.

 A method to find the tadpole term for low-lying surface phonons was developed by Khodel \cite{Khod} and is described in detail
in \cite{KhS}. It is  equal to  \beq
\delta\Sigma^{\rm tad}=\int \frac {d\omega} {2\pi i} \delta_L {g_L} D_L(\omega),\label{tad} \eeq where
$\delta_L {g_L}$ can be found by variation of Eq. (\ref{g_L}) in the field of the
$L$-phonon:
\bea  \delta_L {g_L}&=&\delta_L {\cal F}
A_{\rm ph}(\omega_L){ g_L} + {\cal F} \delta_L A_{\rm ph}(\omega_L){ g_L} \nonumber \\
&+& {\cal F} A_{\rm ph}(\omega_L)\delta_L{g_L}. \label{dgL} \eea The phonon
$D$-function appears in Eq. (\ref{tad}) after connecting  two wavy
$L$-phonon ends in Eq. (\ref{dgL}). This corresponds to averaging of the
product of two boson (phonon) operators $B_L^+B_L$ over the ground
state of the nucleus with no phonons.

\begin{figure}
\centerline {\includegraphics [width=80mm]{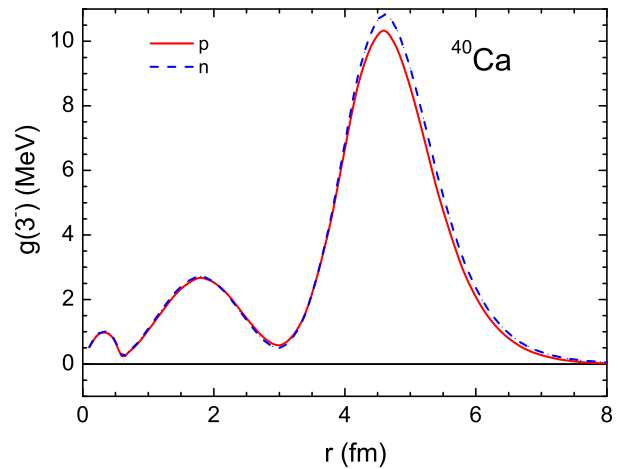}} \caption{ The vertex $g_L$ for the
$3^-_1$ state in $^{40}$Ca.} \label{fig:Ca3-}
\end{figure}

Following Ref. \cite{levels},  we use an approximate way to solve Eq. (\ref{dgL}) based on the surface dominance in the vertex
$g_L({\bf r})$. Indeed, all the $L$-phonons
we consider are the surface vibrations which belong to  the
Goldstone mode corresponding to the spontaneous breaking of the
translation symmetry in nuclei \cite{Khod,KhS}.
For the ghost phonon, $L{=}1,\omega_1{=}0$, which is the lowest energy member of this mode,
 Eq. (\ref{g_L}), due
to the TFFS self-consistency relation \cite{Fay-Khod}, has the exact
solution \beq g_1 ({\bf r}) = \alpha_1 \frac {d U(r) } {d r}Y_{1M}({\bf n}), \label{g1}\eeq where $\alpha_1=
 1 /\sqrt{2\omega
B_1}$ , $B_1=3m A/4\pi $ is the Bohr--Mottelson (BM) mass coefficient \cite{BM2} and
$U(r)$ is the central part of the mean-field potential generated
by the energy functional.

In the general case, the coordinate form of
the amplitude  $g_L({\bf r})$ is very close to that of the ghost phonon:
\beq g_L(r)=\alpha_L \frac {dU} {dr} +\chi_L(r), \label{gLonr}\eeq
where the in-volume correction $\chi_L(r)$ is rather small. The first,
surface term on the right-hand sight. of Eq. (\ref{gLonr}) corresponds to the
BM model for the surface vibrations \cite{BM2}, the amplitude
$\alpha_L$ being related to the dimensionless BM amplitude $\beta_L$
as follows: $\alpha_L=R \beta_L/\sqrt{2L+1}$, where $R=r_0 A^{1/3}$ is the
nucleus radius, and $r_0=1.2\;$fm.

Fig. 3 demonstrates the smallness of the in-volume term of $g_L({\bf r})$ in the case
of the $3^-_1$-state in $^{208}$Pb which is the most collective state in this nucleus.
In more light nuclei, as $^{40,48}$Ca, the surface dominance is not so pronounced but also persists. It is shown in Fig. \ref{fig:Ca3-} for the
$3^-_1$ state in $^{40}$Ca.
If one neglects in-volume contributions, the tadpole PC term (\ref{tad}) can be reduced
to a simple form:
\beq
\delta\Sigma^{\rm tad}_L = \frac {\alpha_L ^2} 2 \frac {2L+1} 3
\triangle U(r). \label{tad-L}\eeq  It should be noted that this relation for the ghost phonon is exact.
Below we neglect the in-volume corrections for all nuclei
considered. To find the phonon amplitudes $\alpha_L$,
 we used the following definition \beq \alpha_L^{\tau}= \frac {g_L^{\tau,{\rm max}}}
  {\left(\frac {dU} {dr}\right)^{\tau,{\rm max}} }, \label{alpL}\eeq with obvious notation.

 \begin{figure}
\centerline {\includegraphics [width=12mm]{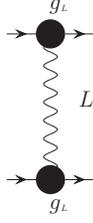}} \caption{The
phonon induced interaction.} \label{fig:PC1}
\end{figure}

\begin{figure}
\centerline {\includegraphics [width=50mm]{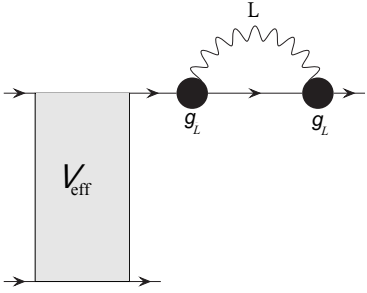}} \caption{An example of the PC ``end'' correction.} \label{fig:PC2}
\end{figure}

 Note that the above scheme for the ghost $L=1$ phonon results in an explicit expression for the ``recoil effect''.
 Details can be found in \cite{levels}.

Let us go to PC corrections to the r.h.s. of Eq. (\ref{eqchi0}). They include the phonon induced interaction, Fig. \ref{fig:PC1},
and the ``end corrections''. An example of them is given in Fig. \ref{fig:PC2}. Partial summing of such diagrams results in the ``renormalization'' of ends:
\beq |\lambda\rangle \to |\widetilde{\lambda}\rangle =\sqrt{Z_{\lambda}^{\rm PC}} |\lambda\rangle. \label{ren_end}  \eeq
In the result, we get
\bea \langle 1 1'|\widetilde{\cal V}_{\rm eff}|22'\rangle &=& \sqrt{Z_1^{\rm PC}Z_{1'}^{\rm PC}Z_2^{\rm PC}Z_{2'}^{\rm PC}}
\nonumber \\
&\times& \langle 1 1'|{\cal V}_{\rm eff} +  {\cal V}_{\rm ind}|22'\rangle. \label{Vtild}\eea

Remind that we deal with the channel with $I{=0},S{=0},L{=0}$. Hence, the states $i,i'$ in (\ref{Vtild}) possess the same single-particle angular momenta, $j_1{=}j_{1'},l_1{=}l_{1'};j_2{=}j_{2'},l_2{=}l_{2'}$. In this case, the explicit expression of the matrix element of ${\cal V}_{\rm ind}$ is as follows:
\bea \langle 1 1'|{\cal V}_{\rm ind}|22'\rangle = - \frac {2 \omega_L} {\sqrt{(2j_1+1)(2j_2+1)}}  \nonumber \\
\times
\frac {\bigl(\langle j_1 l_1|| Y_L || j_1 l_1\rangle (g_L)_{11'}\bigr) \bigl(\langle j_2 l_2|| Y_L || j_2 l_2\rangle (g_L)_{22'}\bigr)^* }{\omega_L^2-(\eps_2-\eps_1)^2
}, \label{Vind} \eea
where $\langle \;|| Y_L || \;\rangle$
stands for the reduced matrix element \cite{BM1}, and $(g_L)_{ii'}$
are the radial matrix elements of the vertex $g_L(r)$. For brevity, we show here explicitly the contribution of the main term of Eq. (\ref{gLS01}) only, with $S{=}0$, $T_{LL0}=Y_{LM}({\bf n})\times \hat 1$. In actual calculations, the component $S{=}1$ is also taken into account but its contribution is always small.

\begin{table}[]
\caption{Characteristics of the low-lying phonons in magic nuclei,
$\omega_L$ (MeV) and $B(EL,{\rm up)}$(${\rm e^2 fm}^{2L}$). \label{tab:BEL}}
\begin{tabular}{c c c c c  }

\noalign{\smallskip}\hline\hline\noalign{\smallskip} $L^\pi$  &
$\omega_L^{\rm th}$   & $\omega_L^{\rm exp}$
&  $B(EL)^{\rm th}$ &  $B(EL)^{\rm exp}$  \\
\noalign{\smallskip}\hline\noalign{\smallskip}

&&    $\qquad\qquad ^{40}$Ca          &&\\
\noalign{\smallskip}
$3^-$    & 3.335     &  3.73669 (5)       &  $1.52 \times 10^4$                  & $1.24   \times 10^4 $ \\
\noalign{\smallskip}\hline\noalign{\smallskip}
&&    $\qquad\qquad ^{48}$Ca          &&\\
\noalign{\smallskip}
$2^+$    & 3.576     &  3.83172 (6)       &  $0.55   \times 10^2 $    & $0.86   \times 10^2 $ \\

$3^-$    & 4.924     &  4.50678 (5)        &$5.701   \times 10^3 $    & $0.67   \times 10^4 $ \\
\noalign{\smallskip}\hline\noalign{\smallskip}
&&    $\qquad\qquad ^{56}$Ni          &&\\
\noalign{\smallskip}
$2^+$    & 2.826     &   2.7006 (7)        & $5.725 \times 10^2 $                   &  \\

$3^-$    & 8.108     &   4.932 (3)         &$2.068   \times 10^4 $    & \\
\noalign{\smallskip}\hline\noalign{\smallskip}
&&    $\qquad\qquad ^{78}$Ni          &&\\
\noalign{\smallskip}
$2^+$    & 3.238     & -         & $3.309 \times 10^2$      & \\

$3^-$    & 6.378     & -         &$1.549   \times 10^4 $    & \\
\noalign{\smallskip}\hline\noalign{\smallskip}
&&    $\qquad\qquad ^{100}$Sn          &&\\
\noalign{\smallskip}
$2^+$    & 3.978     & -    & $1.375\times 10^3$            & \\

$3^-$    & 5.621     & -         &$1.24   \times 10^5 $    & \\

\noalign{\smallskip}\hline\noalign{\smallskip}
&&    $\qquad\qquad ^{132}$Sn          &&\\
\noalign{\smallskip}
$2^+$    & 4.327      & 4.04120 (15)    & $0.104\times 10^4$    & $0.11 (0.03)\times 10^4$\\

$3^-$    & 4.572      & 4.35194 (14)         &$1.29   \times 10^5 $    & \\

\noalign{\smallskip}\hline\noalign{\smallskip}
&&    $\qquad\qquad ^{208}$Pb          &&\\
\noalign{\smallskip}
$3^-$      & 2.684    &  2.615       &$7.093\times 10^5 $    & $6.12 \times 10^5 $\\
$5^-_1$    & 3.353    &  3.198     &$3.003\times 10^8 $    & $4.47 \times 10^8 $\\
$5^-_2$    & 3.787    &  3.708     &$1.785 \times 10^8 $   & $2.41 \times 10^8 $\\
$2^+_1$    & 4.747    &  4.086     &$1.886 \times 10^3 $   & $3.18 \times 10^3 $\\
$2^+_2$    & 5.004    &  4.928     &$1.148 \times 10^3 $   & - \\
$4^+_1$    & 4.716    &  4.324     &$3.007 \times 10^6 $   & - \\
$4^+_2$    & 5.367    &  4.911(?)  &$8.462 \times 10^6 $   & - \\
$6^+_1$    & 4.735    &   -        &$6.082 \times 10^9 $   & - \\
$6^+_2$    & 5.429    &   -        &$1.744 \times 10^{10}$ & - \\

\noalign{\smallskip}\hline\hline\noalign{\smallskip}

\end{tabular}
\end{table}

\begin{table*} \caption{Different PC corrections to odd-even double mass  differences of magic
nuclei.\label{tab:delPCD2pm}}
\begin{center}
\begin{tabular}{ccccc cccl}
\hline\hline\noalign{\smallskip}
&   & $D_2^{(0)}$  & $\delta D_2 (Z^{\rm PC})$
  & $\delta D_2 ({\cal V}_{\rm ind}^{\rm PC})$    &  $\delta D_2(\delta \eps^{\rm PC})$ &\; $\delta D_2^{\rm PC}\;$ &\; $D_2^{\rm PC}$\;\; & \;\; $D_2^{\rm exp}$   \\
\hline

$^{40}$Ca-pp  & $D_2^-$ &     3.001&      -0.427&   -0.539     &  -0.206  & -0.866  &      2.135&     1.94683(19) \\
              & $D_2^+$ &    -2.718&     0.399  &    0.548     &   0.158  &  0.818  &     -1.900&    -2.66622(28) \\

$^{40}$Ca-nn  & $D_2^-$&     4.064&   -0.911  &    -0.971 &   -0.357  &  -1.454     &   2.610&     2.3395(9) \\
              & $D_2^+$&    -3.836&    0.933  &     0.998  &   0.292  &   1.461     &  -2.375&    -3.11785(15) \\

$^{48}$Ca-pp  & $D_2^-$&     2.738&   -0.762 &    0.071  &   0.184   &  -0.663  &    2.075&     2.592(40) \\
              & $D_2^+$&    -3.047&   0.908  &   -0.396  &   -0.280  &  0.449   &   -2.598&    -2.5333(38) \\

$^{48}$Ca-nn  & $D_2^-$&     3.079&   -0.589 &    0.004 &   0.943   & -0.286  &     2.793&     2.6763(23) \\
              & $D_2^+$&    -1.715&   0.344  &    0.096 &  -0.833   &  0.210  &    -1.505&    -1.2141(16)\\

$^{56}$Ni-pp  & $D_2^-$&        2.679&   -0.577  & 0.245    & 0.148  & -0.376 & 2.303&  2.1022(4)   \\
              & $D_2^+$&       -1.461&   0.466   & -0.512   & -0.133 & 0.142  &-1.319& -1.590(50)   \\

$^{56}$Ni-nn  & $D_2^-$&       3.092& -1.035  &  1.271  &  0.197 & -0.787 &   2.305&  2.5517(8)\\
              & $D_2^+$&      -1.931&  0.617  &  -2.754 & -0.201 &  0.413 &  -1.518& -1.9687(7)\\

$^{78}$Ni-pp  & $D_2^-$&     4.161&  -1.913 &    0.619    &  0.343   &  -1.558    & 2.603&  - \\
              & $D_2^+$&    -3.525&  1.873  &   -1.133    &  -0.120  &   1.415    &-2.110&  -1.980(980)\# \\

$^{78}$Ni-nn  & $D_2^-$&    2.330&   -0.614 &   0.427  &    0.116   & -0.221 &   2.109 &  2.240(1190)\#    \\
              & $D_2^+$&   -1.373&  0.365   &  -0.305  &   -0.179   & -0.012  &  -1.385&  - \\

$^{100}$Sn-pp  & $D_2^-$&     2.209&    -0.595  &   0.338  &  0.035  & -0.282      &1.927 & 2.170(410)\# \\
               & $D_2^+$&    -1.190&     0.306  &   -0.133 &  0.022  & 0.188       &-1.002&        -     \\

$^{100}$Sn-nn  &$D_2^-$&     2.651&    -0.737   &   0.363  & 0.075  &  -0.418    & 2.233&  -   \\
               &$D_2^+$&    -1.652&     0.462   &  -0.138  & -0.011 &  0.331    & -1.321& -1.610(540) \\

$^{132}$Sn-pp  & $D_2^-$&     3.184&  -1.506  & -0.015   & -0.982  & -1.198      &1.986    &    2.027(160)\\
               & $D_2^+$&    -2.763&   1.319  & -0.250   & 1.710   & 1.494       &   -1.269&   -1.234(6) \\

 $^{132}$Sn-nn  & $D_2^-$&    2.301&  -0.396   &  0.369   &  -0.009   & -0.161   & 2.140&       2.132(9)\\
                & $D_2^+$&   -1.165&  0.217    &  -0.102  &  -0.045   &  0.094   & -1.071&     -1.227(6)    \\

$^{208}$Pb-pp  & $D_2^-$&   1.680&   -0.824   &   -0.083  &   0.569  & -0.745    &         0.935&    0.627(22)\\
               & $D_2^-$&  -2.286&  1.049     &  -0.167   &  -0.329  &  0.830   &        -1.456&     -1.1845(11)  \\

$^{208}$Pb-nn  & $D_2^-$&   0.778&  -0.275   &   0.174  &   0.205   & -0.113   &    0.665&       0.63009(11)\\
               & $D_2^-$&  -1.156&   0.443   &  -0.691  &  -0.021   & 0.165    &   -0.991&      -1.2478(17)\\
\hline\hline

\end{tabular}
\end{center}
\end{table*}

\begin{table}
\caption{PC contributions to the SP characterisics of  ground states $\lambda_0$  of odd neighbors of magic nuclei. }

\begin{tabular}{ccccc}
\noalign{\smallskip}\hline\hline\noalign{\smallskip}

nucleus  & $\lambda_0$       &  $\eps^{(0)}_{\lambda_0}$ & $\delta \eps_{\lambda_0}^{\rm PC}$ & $Z^{\rm PC}_{\lambda_0}$ \\
\noalign{\smallskip}\hline\noalign{\smallskip} \noalign{\smallskip}

$^{40}$Ca & $1f_{7/2}^p$    &   -2.678  &   0.479  &  0.960 \\
         & $1d_{3/2}^{-p}$ &   -7.265  &   0.122  &   0.966 \\

         & $1f_{7/2}^n$    &    -9.593  &    0.270  &   0.947 \\
         & $1d_{3/2}^{-n}$ &   -14.257  &    0.076  &   0.965 \\

$^{48}$Ca & $1f_{7/2}^p$    &   -9.909 &   -0.031 &    0.899 \\
         & $2s_{1/2}^{-p}$ &   -15.098 &    0.575 &    0.916 \\

         & $2p_{3/2}^n$    &    -5.784 &   -0.062 &    0.940 \\
         & $1f_{7/2}^{-n}$ &    -9.488 &    0.357 &    0.966 \\

$^{56}$Ni & $2p_{3/2}^p$    &    -1.905 &   -0.151 &    0.913 \\
         & $1f_{7/2}^{-p}$ &    -6.276 &    0.530 &    0.963 \\
         & $2p_{3/2}^n$    &   -11.064 &   -0.074 &    0.934 \\
         & $1f_{7/2}^{-n}$ &   -15.588 &    0.486 &    0.945 \\

$^{78}$Ni & $2p_{3/2}^p$    &   -15.526 &   -0.154 &    0.882 \\
         & $1f_{7/2}^{-p}$ &   -20.245 &    0.491 &    0.943 \\
         & $2d_{5/2}^n$    &    -1.477 &   -0.137 &    0.916 \\
         & $1g_{9/2}^{-n}$ &    -5.481 &    0.460 &    0.918 \\

$^{100}$Sn & $2d_{5/2}^p$    &    2.812  &   -0.214 &    0.910 \\
          & $1g_{9/2}^{-p}$ &    -2.345 &    0.492 &    0.939 \\
          & $2d_{5/2}^n$    &   -11.180 &   -0.194 &    0.901 \\
          & $1g_{9/2}^{-n}$ &   -16.449 &    0.511 &    0.939 \\

$^{132}$Sn & $1g_{7/2}^p$    &    -9.892 &    0.227 &    0.967 \\
          & $1g_{9/2}^{-p}$ &   -14.842 &    0.363 &    0.963 \\
          & $2f_{7/2}^n$    &    -2.319 &   -0.131 &    0.939 \\
          & $1g_{9/2}^{-n}$ &    -7.472 &    0.376 &    0.948 \\

$^{208}$Pb & $1h_{9/2}^p$    &    -4.232 &    0.273 &    0.958 \\
          & $3s_{1/2}^{-p}$ &    -7.611 &   -0.023 &    0.930 \\
          & $2g_{9/2}^n$    &    -3.674 &   -0.251 &    0.885 \\
          & $3p_{1/2}^{-n}$ &    -7.506 &   -0.043 &    0.928 \\

\hline\hline

\end{tabular}\label{tab:deps-Z}
\end{table}

\begin{table*} \caption{Double odd-even mass  differences of magic nuclei.\label{tab:D2pm}}
\begin{center}
\begin{tabular}{cccc cccl}
\hline\hline\noalign{\smallskip}
&   & $D_2^{(0)}$   & $D_2(\gamma{=}0.06)$   & $D_2^{\rm PC}$
           &  $D_2^{\rm PC}(\gamma{=}0.06)$ &  $D_2^{\rm PC}(\gamma{=}0.03)$ &\;\;\;$D_2^{\rm exp}$ \\
\hline\noalign{\smallskip}

$^{40}$Ca-pp  & $D_2^-$   &  3.001 &  2.391   &  2.424 & 2.038  &   2.217 &   1.94683(19)\\
              & $D_2^+$   & -2.718 & -2.154   &  -2.202& -1.786 &  -1.987 &  -2.66622(28)\\

$^{40}$Ca-nn  & $D_2^-$&   4.064&     2.955&        2.610&       2.164&  2.153 &    2.3395(9)\\
              & $D_2^+$&  -3.836&    -2.773&       -2.375&      -1.959& -2.148 &   -3.11785(15)\\

 $^{48}$Ca-pp  & $D_2^-$&   2.738&     2.109&          2.075&       1.708&   1.879 &    2.592(40)  \\
               & $D_2^+$&  -3.047&    -2.394&         -2.598&      -2.203&  -2.388 &   -2.5333(38)\\

$^{48}$Ca-nn  & $D_2^-$&   3.079&     2.441&             2.793&      2.282&    2.518&    2.6763(23)   \\
              & $D_2^+$&  -1.715&    -1.335&            -1.505&     -1.229&   -1.353&   -1.2141(16) \\

$^{56}$Ni-pp  & $D_2^-$&   2.679&     2.097&           2.303&         1.892&        2.087&  2.1022(4)       \\
              & $D_2^+$&  -1.461&    -1.107&          -1.319&        -1.098&       -1.203& -1.590(50)       \\

$^{56}$Ni-nn  & $D_2^-$&       3.092&     2.423&           2.305&         1.959&        2.124&   2.5517(8)  \\
              & $D_2^+$&      -1.931&    -1.484&          -1.518&        -1.278&       -1.393&  -1.9687(7)  \\

$^{78}$Ni-pp  & $D_2^-$&       4.161&    2.835&              2.603&       2.120&      2.341&  -  \\
              & $D_2^+$&      -3.525&   -2.213&             -2.110&      -1.687&     -1.878&  -1.980(980)\#\\

$^{78}$Ni-nn  & $D_2^-$&       2.330&   1.815&               2.109&        1.764&       1.927& 2.240(1190)\#      \\
              & $D_2^+$&      -1.373&  -1.111&              -1.385&       -1.231&      -1.302& - \\

$^{100}$Sn-pp & $D_2^-$&       2.209&    1.710&          1.927&         1.599&    1.754&   2.170(410)\# \\
              & $D_2^+$&      -1.190&   -0.869&         -1.002&        -0.824&   -0.907&  -           \\

$^{100}$Sn-nn  & $D_2^-$&    2.651&    2.032&          2.233&         1.837&          2.023& -   \\
               & $D_2^+$&   -1.652&   -1.212&         -1.321&        -1.080&         -1.191& -1.610(540)  \\

$^{132}$Sn-pp  & $D_2^-$&    3.184&    2.281&         1.986&         1.812&          1.905&     2.027(160) \\
               &$D_2^+$&   -2.763&   -1.875&        -1.269&        -1.444&         -1.351&      -1.234(6)  \\

$^{132}$Sn-nn  & $D_2^-$&   2.301&    1.742&          2.140&         1.692&           1.901&    2.132(9) \\
               & $D_2^+$&  -1.165&   -0.900&          -1.071&        -0.879&          -0.967&   -1.227(6)     \\

$^{208}$Pb-pp  &$D_2^-$&   1.680&    1.000&          0.935&         0.718&          0.815&   0.627(22)   \\
               &$D_2^+$&  -2.286&   -1.467&        -1.456&        -1.120&          -1.276&    -1.1845(11)     \\

$^{208}$Pb-nn  & $D_2^-$&   0.778&    0.530&           0.665&        0.494&          0.570&   0.63009(11)  \\
               & $D_2^+$&  -1.156&   -0.821&          -0.991&       -0.820&         -0.899&   -1.2478(17)  \\

\noalign{\smallskip}\hline\hline

\end{tabular}
\end{center}
\end{table*}

\begin{table*} \caption{Difference between theoretical and experimental values of DMD for different versions of the theory.\label{tab:delTHEORY}}
\begin{center}
\begin{tabular}{ccccccl}
\hline\hline\noalign{\smallskip}
& $\gamma{=}0$   & $\gamma{=}0.06$ &  $(\gamma{=}0)^{\rm PC}$  & $(\gamma{=}0.06)^{\rm PC}$ & $(\gamma{=}0.03)^{\rm PC}$  &\;\;\;$D_2^{\rm exp}$ \\
\hline\noalign{\smallskip}

$^{40}$Ca-pp  &  1.054&     0.444&     0.477&     0.091&     0.270&     1.94683(19)\\
              & -0.052&     0.512&     0.464&     0.880&     0.679&    -2.66622(28)\\

$^{40}$Ca-nn  &  1.724&     0.615&     0.270&    -0.175&    -0.187&     2.3395(9)\\
              & -0.718&     0.345&     0.743&     1.159&     0.970&    -3.11785(15)\\

$^{48}$Ca-pp  &   0.146&    -0.483&    -0.517&    -0.884&    -0.713&     2.592(40)\\
              &  -0.514&     0.139&    -0.065&     0.330&     0.145&    -2.5333(38)\\

$^{48}$Ca-nn  &   0.403&    -0.235&     0.117&    -0.394&    -0.158&     2.6763(23)\\
              &  -0.501&    -0.121&    -0.291&    -0.015&    -0.139&    -1.2141(16)\\

$^{56}$Ni-pp  &  0.577&    -0.005&     0.201&    -0.210&    -0.015&     2.1022(4)\\
              &  0.129&     0.483&     0.271&     0.492&     0.387&    -1.590(50)\\

$^{56}$Ni-nn  &   0.540&    -0.129&    -0.247&    -0.593&    -0.428&     2.5517(8)\\
              &   0.038&     0.485&     0.451&     0.691&     0.576&    -1.9687(7)\\

$^{132}$Sn-pp  & -1.529&    -0.641&    -0.035&    -0.210&    -0.117&    -1.234(6)\\

$^{132}$Sn-nn  &  0.169&    -0.390&     0.008&    -0.440&    -0.231&     2.132(9) \\
               &  0.062&     0.327&     0.156&     0.348&     0.260&    -1.227(6)\\

$^{208}$Pb-pp  &  1.053&     0.373&     0.308&     0.091&     0.188&    0.627(22)\\
               & -1.101&    -0.282&    -0.271&     0.065&    -0.091&    -1.1845(11) \\

$^{208}$Pb-nn  &  0.148&    -0.100&     0.035&    -0.136&    -0.060&     0.63009(11) \\
               &  0.092&     0.427&     0.257&     0.428&     0.349&    -1.2478(17)\\

\noalign{\smallskip}\hline\hline

\end{tabular}
\end{center}
\end{table*}

\section{Calculation results}

All calculations are carried out in a self-consistent way with the use of the Fayans EDF with the set DF3-a
of the parameters \cite{Tol-Sap}. We limit ourselves with seven double-magic nuclei, from $^{40}$Ca till $^{208}$Pb. It should be noted
that some of ``new magic nuclei'' are included into consideration just for completeness as corresponding experimental DMD are not known.
Moreover, some nuclei necessary to find  corresponding DMD from Eqs. (\ref{d2npl})--(\ref{d2pmi}) do not exist, being absolutely unstable; hence there is no hope that the corresponding experimental data will appear in future. This is so, e.g. with the $^{98}$Sn nucleus which is a term of the DMD $D_2^{n-}(^{100}{\rm Sn})$ or $^{101}$Sb and $^{102}$Te nuclei which are necessary to find the DMD $D_2^{n-}(^{100}{\rm Sn})$.
Characteristics of the low-lying collective states in these nuclei  are presented in Table \ref{tab:BEL}. As one can see, the
overall agreement of $\omega_L$ and $B(EL)$ values with known experimental data looks reasonable.

As it is well known, see e.g. \cite{levels}, PC corrections are important mainly for single-particle states  close to the Fermi surface. In practice, we solve the ``PC corrected'' equation (\ref{eqPCchi0}) limiting ourselves with two shells nearby the Fermi level. In addition, we as a rule include into the calculation scheme the single-particle  states of negative energy only.
In Table \ref{tab:delPCD2pm}, the effect of each PC correction to each DMD value is given separately. In this set of calculations we put $\gamma{=}0$ in Eq. (\ref{Vef1}) which determines the EPI of the semi-microscopic model, hence $D_2^{(0)}$ means the ``ab initio'' prediction for the DMD. The next columns present separate PC corrections to this quantity. So, the 2-nd column shows the result of
application of Eq. (\ref{Vtild}) with ${\cal V}_{\rm ind}{=}0$, whereas the 3-rd one presents the effect of ${\cal V}_{\rm ind}$ itself with $Z_1^{\rm PC}{=}...{=}Z_{2'}^{\rm PC}{=}1$. The   column 4 shows the effect of PC corrections to the SP energies in Eq. (\ref{eqPCchi0}) only. At last, column 5 presents the total PC effect $\delta D_2^{\rm PC}{=}D_2^{\rm PC}-D_2^{(0)}$, where $D_2^{\rm PC}$ (column 6) is the solution of Eq. (\ref{eqPCchi0}) with all PC corrections included. As it  should be, the value of $\delta D_2^{\rm PC}$ does not equal to the sum of the values in previous three columns because of an interference between different PC effects. Experimental DMD values are found from the mass table \cite{mass}.

General impression from the analysis of Table \ref{tab:delPCD2pm} is that different PC corrections to DMD values are very non-regular, strongly depending on the nucleus under consideration and the two-particle channel as well. The $Z$-factor effect (column 2)  always has the sign opposite to that of $D_2^{(0)}$ value thus suppressing the absolute value of $D_2^{(0)}$. This is a trivial consequence of the $Z^{\rm PC}<1$ condition.  The scale of suppression varies from
$\simeq 15$\% (protons in $^{40}$Ca) to $\simeq 50$\% (protons in $^{132}$Sn or $^{208}$Pb). The suppression value $\delta D_2 ({\cal V}_{\rm ind}^{\rm PC})$ is of the order of the product of $D_2^{(0)}(1{-}Z^{\rm PC}_{\lambda_0})^2$, where the $Z^{\rm PC}_{\lambda_0}$ value is given in Table \ref{tab:deps-Z}. Here $\lambda_0$ denotes the single-particle state of a nucleon added to (or removed from) the magic nucleus under consideration. These two quantities should coincide if we use the ``diagonal approximation'' retaining in Eq. (\ref{eqPCchi0}) the term $\lambda_1=\lambda_2=\lambda_3=\lambda_4=\lambda_0$ only. However, non-diagonal terms play some role in this equation making these two quantities equal only approximately.

The sign of the PC effect due to the induced interaction in the major  part cases coincides with that of $D_2^{(0)}$, i.e. corresponds to an additional attraction. However, there are exceptions, e.g. $^{40}$Ca nucleus, both proton and neutron modes. As a rule, the value of this effect is less than that of the $Z$-factors, but there are cases where it is rather big. This is so, e.g. in both neutron modes in $^{40}$Ca. It occurs due to appearance of a small denominators in Eq. (\ref{Vind}) corresponding to the states $\lambda_2{=}1f_{7/2}$ and $\lambda_1{=}1d_{2/2};2s_{1/2}$. This effect is even stronger
in the neutron $D_2^+$ mode in $^{56}$Ni due to the same SP states leading to small denominators in Eq. (\ref{Vind}). Such cases of anomalously large value of the PC correction due to the induced interaction is a signal that the $g_L^2$ perturbation theory does not work sometimes even in magic nuclei and higher order $g_L^2$ terms should be taken into account. The use of the PC corrected single-particle energies in Eq. (\ref{Vind}) is one of possible ways. Fortunately, this term in Eq. (\ref{Vtild}) for the neutron $D_2^+$ mode in $^{56}$Ni is strongly suppressed with the $Z$-factors so that the resulting PC effect (column 5) turns out to be moderate. However, we are forced to interpret this result,
just as those for the neutron modes in $^{40}$Ca, as very approximate.

At last, go to the single-particle energy effect (column 4). In the ``diagonal approximation'' it should be equal to the double value of $\delta \eps^{\rm PC}_{\lambda_0}$, see Table \ref{tab:deps-Z}. As above $\lambda_0$ is the single-particle state of the odd nucleon, added to or removed from the double magic core. As it can be seen in the table, this quantity varies strongly depending on the nucleus and the state $\lambda_0$. However, again there is no complete coincidence between $\delta D_2(\delta \eps^{\rm PC})$ and $\delta \eps^{\rm PC}_{\lambda_0}$ values due to some effect of non-diagonal terms in Eq. (\ref{eqPCchi0}). Moreover, sometimes these two quantities even have opposite signs, but always they are of the same order of magnitude. We did not show contributions to  $\delta \eps^{\rm PC}_{\lambda_0}$ of the pole and tadpole diagrams separately. It can be found in \cite{levels} where it is shown that the tadpole term, as a rule, diminishes the value of $|\delta \eps^{\rm PC}|$ at approximately $30-50$\%. Partially due to this compensation, the single-particle energy effect is, as a rule, significantly less than two PC effects discussed previously. However, there is a case, both neutron modes in $^{48}$Ca, where this PC effect dominates. Thus, all three PC effects under consideration should be taken into account on equal footing.

On average, account for PC effects makes agreement with experiment better, often significantly. However, there are several cases, the proton $D_2^+$ mode in $^{40}$Ca and $^{56}$Ni and the  neutron $D_2^+$ mode in $^{56}$Ni, where they make agreement worse.

In Table \ref{tab:D2pm} we analyze together the PC effects considered above with the suppression of the EPI in the semi-microscopic model with non-zero value of the phenomenological parameter $\gamma$. Notation is similar to that in Table \ref{tab:delPCD2pm}, i.e. the first two columns of   Tables \ref{tab:delPCD2pm} and \ref{tab:D2pm} coincide. Further, $D_2(\gamma{=}0.06)$ means the solution of Eq. (\ref{eqchi0})
(i.e. that without PC effects) with $\gamma{=}0.06$ in Eq. (\ref{Vef1}). The column 3 of this table coincides with the column 6 of Table \ref{tab:delPCD2pm}. Now, $D_2^{\rm PC}(\gamma{=}0.06)$ (column 4) and $D_2^{\rm PC}(\gamma{=}0.03)$ (column 5) mean  the solutions of Eq. (\ref{eqPCchi0}) with ${\cal V}_{\rm eff}$ in Eq. (\ref{Vtild}) found with $\gamma{=}0.06$ and $\gamma{=}0.03$, correspondingly. At first sight, PC corrections make agreement with the data better and the version of $D_2^{\rm PC}(\gamma{=}0.03)$ looks on average the best one among five theoretical columns. To make the comparison with experiment more transparent, we present in Table \ref{tab:delTHEORY} differences between each of these theoretical prediction and the corresponding experimental value. 18 cases are chosen where the experimental data exist and possess  sufficiently high accuracy. Let us concentrate mainly on comparison of the column of  $(\gamma{=}0.03)^{\rm PC}$  with the one corresponding to $\gamma{=}0.06$ without PC corrections. The latter is a representative of the original semi-microscopic model without PC corrections with the optimal description of the pairing gap \cite{Pankr3,Sap1} and DMD of non-superfluid components of semi-magic nuclei \cite{Gnezd1,Gnezd2,Gnezd3} as well. The situation is essentially different for lighter nuclei, from $^{40}$Ca till $^{56}$Ni, and for heavier ones, beginning from $^{132}$Sn. In the first case, the situation is ``fifty-fifty'', i.e. approximately in a half of the cases PC corrections make agreement better and in a half, worse. Agreement typically becomes worse in the cases discussed above where the applicability of the perturbation theory for the induced interaction is questionable. Especially strong disagreement arises for the neutron $D_2^+$ mode in $^{40}$Ca. Absolutely another situation takes place in the lower part of Table \ref{tab:delTHEORY} for heavy nuclei. Here the PC corrections to DMD values taken into account make agreement better in all cases. Sometimes the improvement is significant, e.g. for the proton $D_2^+$ mode in $^{132}$Sn and $^{208}$Pb nuclei.

\section{Conclusion}
A method is developed to account for  the PC  effects in the problem of finding odd-even DMD of magic
nuclei within the {\it ab initio} approach starting from a realistic $NN$-potential.
Recently, the semi-microscopic model of the EPI ${\cal V}_{\rm eff}$ developed first for the pairing problem \cite{Pankr1,Pankr2,Sap1} was applied to the odd-even DMD  for non-superfluid subsystems of semi-magic nuclei \cite{Gnezd1,Gnezd2,Gnezd3}. The DMD values are found by solving the in-medium Bethe-Salpeter equation with the same EPI ${\cal V}_{\rm eff}$  as that in the pairing gap equation. The semi-microscopic  model starts from the interaction ${\cal V}_{\rm eff}$ found in terms of a free $NN$-potential (Argonne $v_{18}$ in our case), the gap equation being solved in the basis with the bare mass $m^*{=}m$. Then the obtained EPI is supplemented with a phenomenological repulsive $\delta$-term proportional to a dimensionless parameter $\gamma$. The value of $\gamma{=}0.06$ found in \cite{Pankr2} to reproduce experimental gap values turned out to be also optimal for describing the DMD values in non-superfluid sub-systems \cite{Gnezd1,Gnezd2,Gnezd3}. The phenomenological addendum supposedly  embodies on average three different corrections to the simple BCS scheme \cite{milan2,milan3}, the
PC contribution, that from the effect of the effective mass $m^*{\neq}m$ and the one due to the high-lying excitations. The last two phenomena are presumably universal and their description with a universal parameter $\gamma$ looks reasonable. On the contrary, low-lying phonon characteristics vary significantly depending on the nucleus under consideration. Therefore, the PC contributions to the gap or the DMD values may fluctuate from nuclei to nuclei.
In this work, we analyze the PC corrections to DMD values found within the semi-microscopic with possible change of the parameter $\gamma$.

 We limit ourselves with seven magic nuclei, from $^{40}$Ca till $^{208}$Pb.  Three PC effects are taken into
account, the phonon induced effective interaction,  the  ``end'' correction, and the change of the
single-particle energies. The perturbation theory in $g^2_L$, where $g_L$ is the vertex of the $L$-phonon creation, is used. However, higher order in $g^2_L$
terms are included in the calculation scheme with partial summation of the end diagram. It results in a renormalization of the end single-particle wave functions, $|\lambda\rangle \to \sqrt{Z_{\lambda}^{\rm PC}}|\lambda\rangle$.    PC  corrections to single-particle energies are found self-consistently
with an approximate account for the tadpole diagram.   For lighter part of the set of magic nuclei, from $^{40}$Ca till $^{56}$Ni, the cases divide approximately in half between those where the PC corrections to DMD values make agreement with experiment better and the ones with the opposite result. In the major part of the ``bad'' cases, a poor applicability of the perturbation theory for the induced interaction, because of appearance of ``dangerous'' terms with small energy denominators, is the most probable reason of the disagreement. For intermediate nuclei, $^{78}$Ni and  $^{100}$Sn, there is no sufficiently accurate data on their masses. Finally, for heavier nuclei, $^{132}$Sn and  $^{208}$Pb, PC corrections to DMD always make agreement with the experiment better. In this case, the optimal value of the phenomenological parameter of the semi-magic model reduces to $\gamma{=}0.03$. This result makes it promising a programme of systematic account for the PC corrections to the semi-microscopic model. There are two possible ways in this direction. The first one is consideration of a wider amount of nuclei, including semi-magic ones, but with more careful separation of cases with good applicability of the perturbation theory in $g^2_L$. The second one is an attempt to develop a more consistent theory with higher in $g^2_L$ terms for consideration of the dangerous terms. Both programs are now in progress.

\acknowledgments The work was partly supported  by the Grant NSh-932.2014.2 of the Russian Ministry
for Science and Education, and by the RFBR Grants 13-02-00085-a, 13-02-12106-ofi\_m,
14-02-00107-a, 14-22-03040-ofi\_m. Calculations were partially carried out on the Computer Center of Kurchatov Institute.
E. S. thanks the INFN, Seczione di Catania, for hospitality.
{}
                \

\end{document}